\documentclass[11pt]{article}
\usepackage{moriond,epsfig}

\def\Journal#1#2#3#4{{#1} {\bf #2}, #3 (#4)}

\def\be{\begin{equation}}
\def\ee{\end{equation}}
\def\bea{\begin{eqnarray}}
\def\eea{\end{eqnarray}}

\begin{document}
\vspace*{4cm}
\title{RAPIDITY GAPS IN GLUON JETS / COLOR RECONNECTION AT LEP}

\author{ M. GIUNTA }

\address{Department of Physics, University of California,\\
Riverside, CA 92521, USA}

\maketitle\abstracts{
Gluon jets with a large gap in the rapidity distribution of particles within the jet are very sensitive to color reconnection effects.
Gluon jets of this type, selected from hadronic $\mathrm{Z^0}$ decay events produced in $\mathrm{e^+e^-}$ annihilations at LEP, are used to test three QCD models which include a simulation of color reconnection: one in the Ariadne Monte Carlo, one in the Herwig Monte Carlo, and the other by Rathsman in the Pythia Monte Carlo. We conclude that color reconnection as implemented by the Rathsman and Ariadne models is disfavored. We don't obtain a definite conclusion concerning the Herwig color reconnection model.}

\section{Introduction}
Rapidity  is a standard variable used to describe the phase space distribution of particles in a multihadronic  event. It is defined by   $y= \frac{1}{2} \ln \left( \frac{E+p_{\parallel}}{E-p_{\parallel}} \right)$ with $E$ the energy of a particle and $p_{\parallel}$ the 3-momentum component along an axis~\footnote{usually the thrust, jet or beam axis}.

\begin{center}
\begin{figure}
\begin{tabular}{cc}
\hspace{2.cm}\includegraphics[width=3.cm]{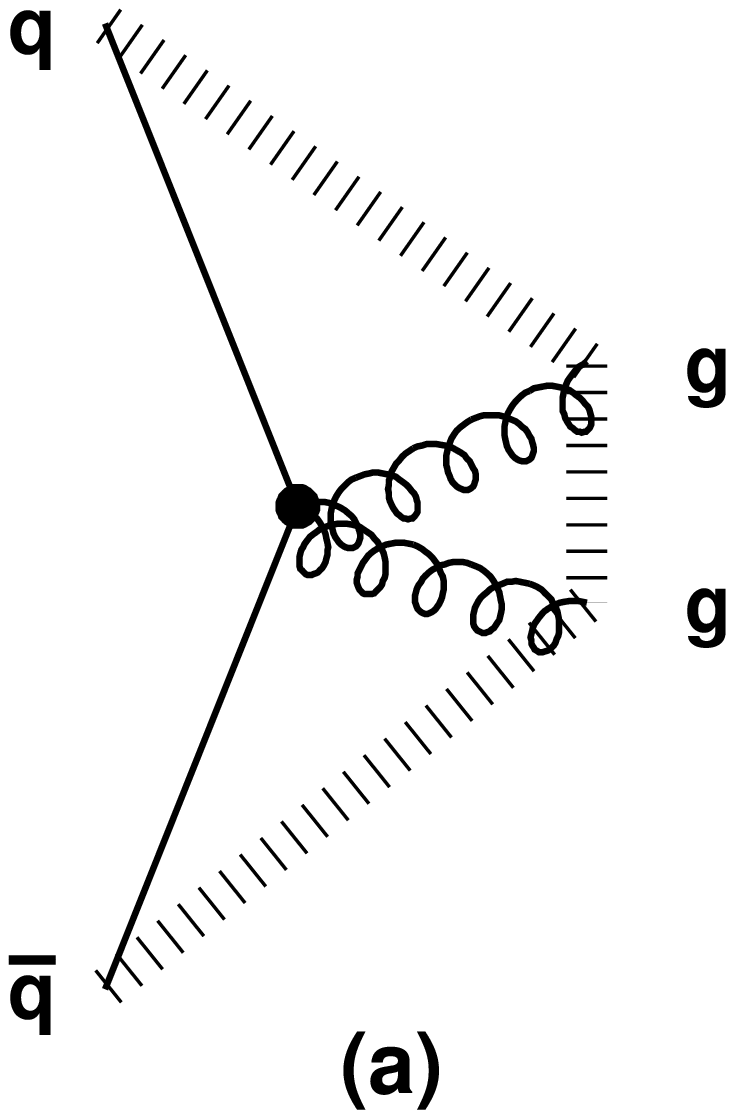}&
\hspace{3.cm}\includegraphics[width=3.cm]{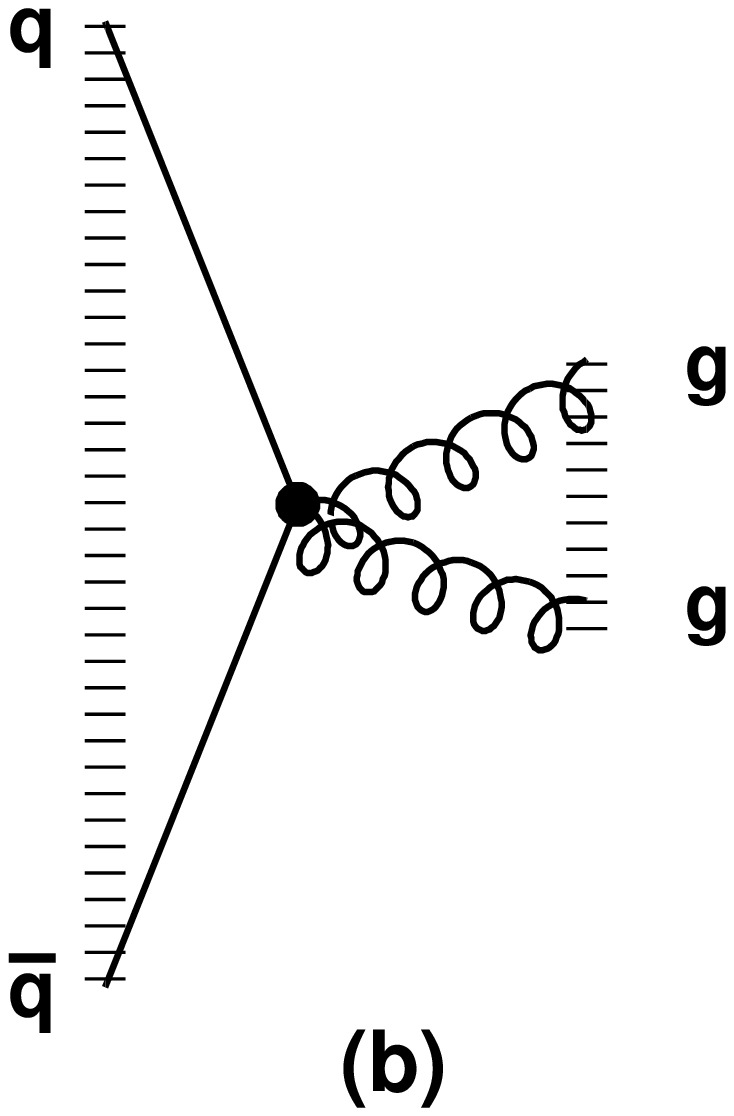}\\
\end{tabular}
\label{fig_diag}
\caption{Schematic illustration of events with (a) standard color flow and (b) reconnection.}
\end{figure}
\end{center}

A rapidity gap event is an event in which two populated regions in rapidity are separated by an empty region.
Color reconnection (CR), i.e  a rearrangement of the color structure of an event, can be a source of this type of events.
In the standard Monte Carlo models only the simplest configuration is allowed (Fig.\ref{fig_diag}a),  where the color flux tube is stretched from a quark to the corresponding antiquark without crossing. If CR is included in the Monte Carlo, we can have  more complex diagrams, in which string segments can either cross or appear as disconnected entities whose endpoints are gluons (Fig.\ref{fig_diag}b); these diagrams are of higher order in Quantum Chromodynamics (QCD) and are suppressed by a factor $1/N_C^2$, where $N_C=3$ is the number of colors.

In events with an isolated gluonic system (Fig.\ref{fig_diag}b) a rapidity gap can form between the particles coming from the hadronization of the isolated segment - often the leading (highest rapidity) part of  a gluon jet - and the rest of the event. Thus rapidity gaps in gluon jets provide a sensitive means to search for color reconnection effects.

\section{Color reconnection study}
To establish the sensitivity of the analysis to processes with color reconnection, multihadronic events are generated  using Monte Carlo simulations, both with and without the effects of color reconnection. 

The models without color reconnection used in the following studies  are the Jetset~\cite{jet}, Herwig~\cite{hw1,hw2} and Ariadne~\cite{ar} Monte Carlo programs, version 7.4, 6.2 and 4.11 respectively. The models which incorporate color reconnection are the model of L\"{o}nnblad~\cite{ar-cr} implemented in the Ariadne Monte Carlo~\footnote{the three different existing implementations, corresponding to settings of the parameter MSTA(35)=1, 2, or 3 are equivalent for hard processes involving a single color singlet system, such as $\mathrm{Z^0}$ decays.}, the color reconnection model~\cite{hw2} in the Herwig Monte Carlo, and a model introduced by Rathsman~\cite{rath} implemented in the Pythia Monte Carlo, version 5.7. In the following I refer to these as the Ariadne-CR, Herwig-CR, and Rathsman-CR models, respectively.

The general strategy used in this type of analysis is the following.
After verifying that all the models, with and without CR, give a good description of the global features of hadronic events, gluon jets with a rapidity gap are selected. This is done requiring a large value for the smallest particle rapidity in a jet, $y_{min}$, or  a large value for the maximum difference between the rapidities of adjacent rapidity-ordered particles, $\Delta y_{max}$. The \textit{leading part} of the jet is defined by the particles beyond the gap. 
Then  the predictions given by the different Monte Carlo models for the distributions of the charged particle multiplicity $n^{ch}_{leading}$ and the total electric charge $Q_{leading}$ of the leading part of the jet are compared to the experimental data.

\subsection{Rapidity gap analysis}

In a recent note~\cite{op-pn518} OPAL chooses a relatively complete and uncorrelated set of distributions sensitive to global event properties and evaluates the total $\chi^2$ value between the hadron level predictions of the different models and the corrected data.
For a total of 81 bins, the $\chi^2$ is 36.9 for Ariadne and 32.4 for Ariadne-CR; 200.7 for Jetset and 243.5 for Rathsman-CR; 127.9 for Herwig and 151.6 for Herwig-CR. This result illustrates the fact that color reconnection has only a small effect on the global features of inclusive $\mathrm{e^+e^-}$ events.

All the events are then forced to three jets using  the Durham~\cite{durham}  jet finder with a variable value for the resolution scale $y_{cut}$ and the two quark ($\mathrm{q~ or ~\bar{q}}$) jets are identified using a b-tagging technique~\cite{b-tag}, the remaining jet is  the gluon jet.

The rapidity gap is defined using charged and neutral particles  and requiring $y_{min}>1.4$,  or $\Delta y_{max}>1.3$ for jets with $y_{min}<1.4$,  where 1.4(1.3) is chosen because it is the value where the prediction for the $y_{min}$ ($\Delta y_{max}$) distribution given by the models with CR starts to deviate from that of the corresponding model without CR.

The $n^{ch}_{leading}$ and $Q_{leading}$ distributions are normalized to the total number of selected jets before the rapidity gap requirement. The results are shown in Fig.2. Both Rathsman-CR and Ariadne-CR predict a large excess of entries  at $n^{ch}_{leading}=2$ and 4 and at $Q_{leading}=0$. These are consequences of events with an isolated gluonic system in the leading part of the gluon jet, which is neutral and decays into an even number of charged particles. Herwig-CR predicts a less striking excess for $3\leq n^{ch}_{leading}\leq 5$. Jetset and Ariadne give predictions 15-20 \% too low for the $Q_{leading}=0$ bin but as there is no spiking behaviour in the data for the $n^{ch}_{leading}$ distribution -- the most unambiguous signal for reconnection --  it cannot be concluded this is due to CR.

\begin{center}
\begin{figure}
\hspace*{2cm}
\includegraphics[width=10.cm]{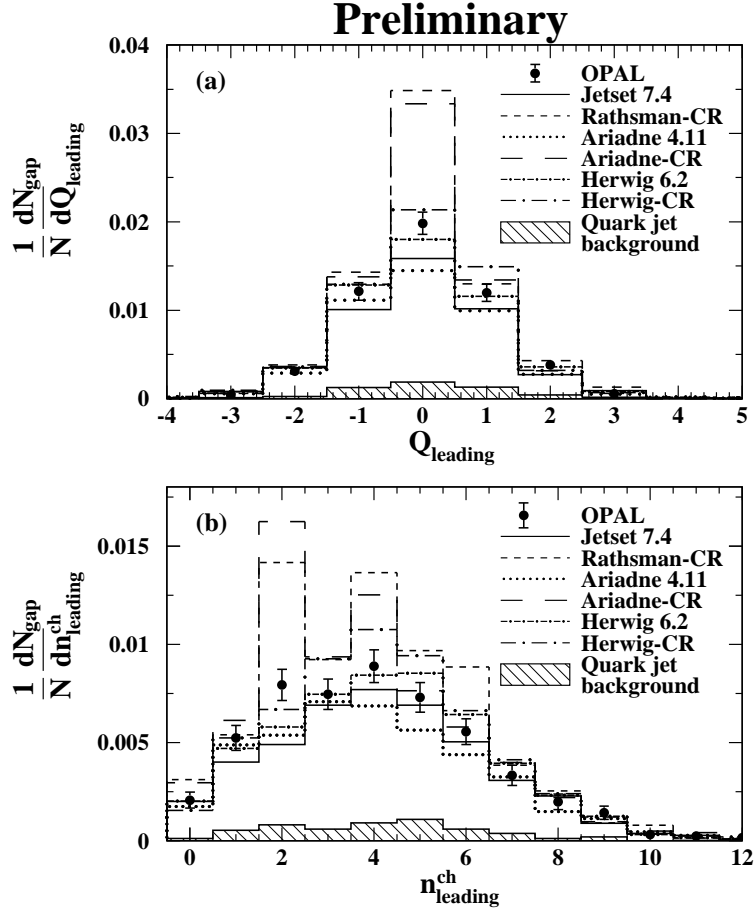}
\label{fig_nch}
\caption{Distributions of (a) $n^{ch}_{leading}$ and (b) $Q_{leading}$ in the leading part of gluon jets. ``N'' represents the total number of selected gluon jets and ``$\mathrm{N_{gap}}$'' the number of gluon jets with a rapidity gap.} 
\end{figure}
\end{center}

ALEPH, performing a similar analysis, observes the same excess in  the $Q_{leading}=0$ bin for Rathsman-CR and Ariadne-CR.

The parameter values used for the different Monte Carlo models (Jetset, Rathsman-CR, Ariadne and Ariadne-CR) are obtained from fits to global quantities.
Three jet events are selected using Durham with fixed $y_{cut}=0.01$, then gluon jets are identified using energy ordering: the jet energies  are recalculated using massless kinematics and ordered in energy, the lowest energy jet is identified as the gluon jet. 

The rapidity gap is defined using charged particles only and requiring $y_{min}=1.5$.
For comparison, quark jets from two jet events and the two higher energy jets from three jet events are also analyzed. The distributions are normalized to the number of jets with a rapidity gap.

It is observed that all the models give an excellent description of the $Q_{leading}$ distribution of quark jets. For the gluon jets, the models with CR predict an excess in  the $Q_{leading}=0$ bin and the models without CR predict too few neutral jets, consistent with the result by OPAL.
ALEPH performs the following checks: defines the gap using $y_{min}$=1.7 or 2.0 or using charged and neutral particles, changes some cuts used to define jets; they reach qualitatively the same conclusions.

\subsection{Models re-tuning}
OPAL then checks if it is possible to vary the Monte Carlo parameters in order to describe the gluon jet data while continuing to give a good description of the global features of the events.

The Rathsman-CR  model describes these two distributions if the value $R_0=0.0085\pm0.0075(stat.)\pm0.0087(syst.)$ is used for the CR suppression factor, this result is consistent with $R_0=0$, that means no CR allowed.

Then OPAL tries to re-tune the models with CR so that they can describe the $n^{ch}_{leading}$ and $Q_{leading}$ data distributions. For Rathsman-CR this is only  possible using $Q_0=5.5~ \mathrm{GeV/c^2}$ and $b=0.27 ~\mathrm{GeV^{-2}}$, but the description of inclusive $\mathrm{Z^0}$  measurements is severely degraded -- the total $\chi^2$ increases from 243.5 to 1117.7 --. For Ariadne-CR  that result is only obtained  with $p_{T,min}=4.7~ \mathrm{GeV/c}$ and $b=0.17 ~\mathrm{GeV^{-2}}$, but also in this case the consequence is a bad description of the global features of hadronic events -- the total $\chi^2$ goes from 32.4 to 3019.3 --.
OPAL concludes that the Rathsman-CR and the Ariadne-CR models are both disfavored.

\section{Summary and Conclusion}
Results by the OPAL and ALEPH collaborations have been presented.

It is observed that models with color reconnection predict a large excess of gluon jets with a rapidity gap and large spikes in some distributions sensitive to color reconnection effects, namely the charged particle multiplicity $n^{ch}_{leading}$ and the total electric charge $Q_{leading}$ of the leading part of the gluon jet, in disagreement with data. 
A tuning of the models with color reconnection in order to describe the $n^{ch}_{leading}$  and  $Q_{leading}$   distributions results in a severe degradation of the description of the global features of the events.

It is concluded that color reconnection as currently implemented in the Rathsman-CR and Ariadne-CR models is strongly disfavored. No definite conclusion is obtained concerning the Herwig-CR model.

\end{document}